\documentclass[aps,prab,groupedaddress,floats,twocolumn]{revtex4-2}

\usepackage[english] {babel}
\usepackage{amsfonts,graphicx,soul,natbib,mathrsfs}

\usepackage{epstopdf}
\usepackage{xcolor}
\usepackage{soul}
\usepackage{physics}
\usepackage{graphics}
\usepackage{textcomp}

\usepackage[colorlinks=true]{hyperref}

\usepackage{dcolumn}
\usepackage{bm}

\begin{document}

\title{Single-shot Transverse Wakefield Mapping with a Hollow Electron Beam}

\author{A. Halavanau}%
\affiliation{SLAC National Accelerator Laboratory, Menlo Park, California 94025, USA}%

\author{P. Piot}%
\affiliation{Department of Physics, Northern Illinois University, DeKalb, IL 60115, USA}%
\affiliation{Argonne National Laboratory, Lemont, IL 60439, USA}%

\author{S. S. Baturin}%
\email{s.s.baturin@gmail.com}%
\affiliation{School of Physics and Engineering,
ITMO University, St. Petersburg, Russia 197101}%

\date{\today}
\begin{abstract}
Beam-driven wakefield accelerators are foreseen to enable compact accelerator-based light sources and play a critical role in future linear-collider concepts. This class of wakefield acceleration has been extensively studied over the last four decades with a focus on demonstrating its ability to support high-accelerating gradient and, most recently, enhanced transformer ratios. Yet, the associated detrimental transverse wakefields have not been examined in as many details due to the limited diagnostics available. In this paper, we introduce a beam-based single-shot transverse-wakefield measurement technique. The approach employs a witness ''hollow" electron beam to probe the wakefields generated by a drive bunch. We show how the transverse distortions of the hollow probe provide a direct measurement of the wakefield distribution within the area circumscribed by the probe. The ability to directly measure a full structure of the transverse wakefield could help to develop mitigation schemes and ultimately suppress the adverse beam-break-up instabilities. We discuss a practical implementation of the method and demonstrate its performance with the help of start-to-end simulations.
\end{abstract}

\maketitle

\section{Introduction}
Collinear beam-driven wakefield field acceleration -- or collinear wakefield acceleration (CWA) -- relies on the deceleration of high-charge ($\mathcal{O}$~[10-100~nC])  ``drive" bunches through slow-wave structures (waveguides or plasmas) to excite electromagnetic wakefields \cite{Weiland:1982,Weiland:1985vg,Wilson:1985,PChen1985,Gai:1988,Rosen:1988}. The produced wakefields are directly employed to accelerate a lagging "witness" bunch. CWAs based on sub-meter-long waveguides and plasmas were experimentally shown to support  ${\mathcal O}$~[GV/m] accelerating fields~\cite{Blumenfeld2007,doi:10.1142/S1793626816300061,OShea2016}. More recently, high-transformer ratios were demonstrated in several experiments~\cite{Gao:2018,Loisch:2018,Roussel:2020}. These achievements open the path toward the design of small-footprint, high-gradient, efficient accelerators. 

Most of the research effort has so far been focused on the acceleration process and associated longitudinal beam dynamics. Yet, taking full advantage of the high acceleration gradient potentially supported by CWA is ultimately limited by the time-dependent transverse forces which are experienced by off-axis particles \cite{Chao}. These transverse wakefields can degrade the transverse emittances of drive-witness pair due to the relative deflection of the head and tail of the bunch. It is especially a significant limitation to the accelerator efficiency as it can strongly affect the decelerating drive bunch and leads to a beam-break-up (BBU) instability where the particle offsets grow exponentially and may even result in most of the drive bunch being lost~\cite{BBU1,BBU2,BBU3}. 

Mitigation of the adverse effects and witness bunch emittance preservation is critical to the practical implementation of GV/m-scale gradient CWA. There are currently two mitigation strategies. The first approach involves the engineering of structures with suppressed dipole wakefield via manipulation of the material properties. The second approach utilizes transverse drive beam shaping techniques. Examples following the latter approach include use of flat or elliptic beams \cite{PhysRevE.56.7204,PhysRevLett.108.244801,Brendan:2020}, dual driver injection \cite{Baturin:dDrive} in a planar waveguide with a retarding material (dielectric, corrugation, etc.). Likewise, mode-filtering technique has been proposed in cylindrical dielectric waveguides~\cite{doi:10.1063/1.348823} and recently studied in the planar geometry with a photonic crystal \cite{woodpile} and Bragg \cite{bragg} loading.
Additionally, the impact of transverse wake on the beam dynamics can be alleviated by proper lattice design and beam control~\cite{BBU2,BBU3}.  

Ultimately, the design of structures capable of suppressing the most harmful components of the transverse wakefields and devising beam dynamics techniques that mitigate their impact will require a precise understanding of the transverse-wakefield distribution. Specifically, a fast single-shot method capable of direct measurement of the transverse-field distribution at various axial positions is very instrumental. It can guide and experimentally validate the mitigation schemes, especially in the cases where theoretical investigations are obscured by the significant complexity of the problem. 

In this paper, we discuss a single shot, beam-based method to reconstruct a snapshot of transverse wakefield at a given delay behind the drive bunch. The proposed technique measures the integrated transverse kick excited by the drive bunch on a closed contour. This information is then used to reconstruct the wakefield in the area enclosed by the contour. In our method, the witness is bunch tailored to a hollow, necklace-like transverse distribution in order to sample the integrated transverse-wakefield kick over a contour.

The proposed wakefield-mapping method is relatively simple and could be implemented at any existing photoinjector e-beam facility with minor hardware modifications, and using conventional beam diagnostics (transverse-density ``screens"). 

As an illustrative example, we present start-to-end simulations of a possible proof-of-principle experiment at the Argonne Wakefield Accelerator (AWA) facility \cite{Power:2010zza}. The method could also be implemented at the FACET-II \cite{FACET} facility. Preliminary studies of a hollow witness beam generation in the LCLS photoinjector are discussed in Ref.\cite{Halavanau:IPAC20}.

\section{Transverse wakefield reconstruction method}

In this section, we discuss the theoretical foundation of the proposed method. We especially detail a reconstruction algorithm and demonstrate its application to retrieve the wake potential associated with a simple geometry modelled by a semi-analytical wake.

\subsection{Theoretical background}
\begin{figure}[t]
    \centering
    \includegraphics[width=0.99\linewidth]{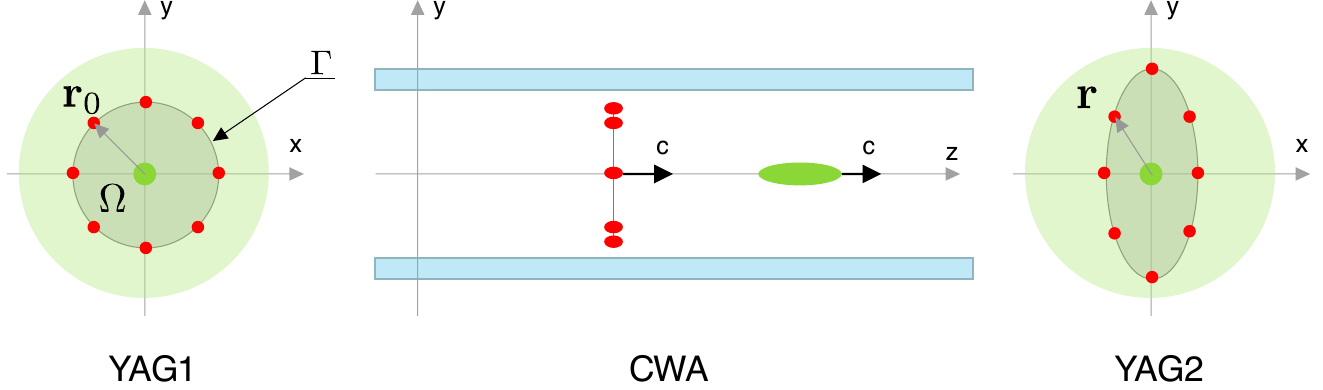}
    \caption{Schematics of the model. Left panel: drive and witness beams projections onto the $xy$-plane, upstream of the CWA section.
    Middle panel: Electron beamlets in a circular or "necklace''-like arrangement (red) are traveling at a certain distance behind the drive beam (green) in a collinear wakefield accelerator (CWA) section (blue).  Right panel: drive and witness beams projections onto the $xy$-plane, downstream of the CWA section. The initial shape of the witness beam is modified according to the transverse wakefield generated by the drive beam.}
    \label{Fig:2}
\end{figure}

We consider a drive bunch injected into a wakefield accelerator section and a witness bunch that consists of a circular (or ''necklace") beamlet arrangement propagating behind the drive bunch with a controllable delay; see Fig.~\ref{Fig:2}. The setup includes a pair of scintillating screens (YAG1 and YAG2) to measure the beam's transverse distribution upstream and downstream of the CWA. We assume both drive and witness bunches to be ultra-relativistic so the speed $V \approx c$ is close to the speed of light as commonly assumed in wakefield accelerators. While propagating through the CWA section, the electromagnetic fields generated by the drive bunch are experienced by the witness bunch. The associated forces yield the witness-bunch shape to change in the longitudinal and transverse directions. The transverse evolution of the drive-witness pair can be characterized by capturing the transverse beam distribution on the YAG2 screen. Under certain conditions, it is possible to infer the average transverse force acting on the witness beamlets by simply comparing the beam distribution measured on the YAG screen located upstream and downstream of the CWA (YAG1 and YAG2 in Fig.~\ref{Fig:2} respectively). The knowledge of the transverse force acting on each beamlet is sufficient to reconstruct the value of this force at any point, inside the contour surrounded by the beamlets. To justify the latter statement let's first consider the wave equation associated with the longitudinal component of the electric field in the vacuum
\begin{align}
\label{eq:EzD}
    \Delta E_z-\frac{\partial^2 E_z}{c^2\partial t^2}=4\pi\frac{\partial \rho}{\partial z}+4\pi\beta \frac{\partial \rho}{c\partial t}.  
\end{align}
Here and further we consider CGS unit system, $\beta=V/c$ is the relativistic beta-factor and $\rho$ - is the charge density. 

By definition, the longitudinal wake potential is connected to the $E_z$ (see for example Ref.\cite{Zotter}) following  
\begin{align}
\label{eq:wkdef}
    W_\parallel(\zeta)=-\frac{1}{Q}\int\limits_{-\infty}^{\infty}dz E_z\left(z,t=\frac{z+\zeta}{c}\right).
\end{align}
Given the connection $ct=z+\zeta$ one can express partial derivative with respect to $z$ and $t$ in the Eq.\eqref{eq:EzD} through the partial derivative with respect to $\zeta$ as 
\begin{align}
\label{eq:Ezm}
    \Delta_\perp E_z+\frac{1}{\gamma^2}\frac{\partial^2 E_z}{\partial \zeta^2}=-\frac{4\pi}{\gamma^2}\frac{\partial \rho}{\partial \zeta}.  
\end{align}
Here $\Delta_{\perp}$ is the transverse component of the Laplace operator and $\gamma\equiv 1/\sqrt{1-\beta^2}$ is the Lorentz factor. For highly energetic beams ($\gamma \gg 1)$ the terms on the order $\mathcal{O}[1/\gamma^2]$ become negligible. This observation, together with Eq.\eqref{eq:wkdef}, leads to the following approximate equation for the longitudinal wake potential
\begin{align}
\label{eq:lWP}
  \Delta_\perp  W_\parallel\approx0.
\end{align}
The latter equation  is exact for the limiting case of $\beta=1$ and is widely considered (see e.g. Refs.\cite{Zotter,Chao,KBane}) as the main theoretical approximation.

The longitudinal wake potential $W_\parallel$ is connected to the transverse wake potential $\mathbf{W}_\perp=(W_x,W_y)^{T}$ through the Panofsky-Wenzel theorem (see Ref.\cite{PW,Henke,Zotter})
\begin{align}
\label{eq:PVT}
    \frac{\partial \mathbf{W}_\perp}{\partial \zeta}=\nabla_\perp W_\parallel. 
\end{align}

We integrate Eq.\eqref{eq:lWP} over $\zeta$, apply $\nabla_\perp$, and accounting for Eq.\eqref{eq:PVT}, arrive at
\begin{align}
\label{eq:TRLP}
     \Delta_\perp W_x\approx0 , \\ \nonumber
     \Delta_\perp W_y\approx0.
\end{align}
These equations indicate that both components of the transverse wake potential are harmonic functions of the transverse coordinates and therefore are completely defined by the value on some closed curve $\Gamma$ within the domain $\Omega$ enclosed by this curve~\cite{Shabat,silverman}.

We now return to the consideration of the witness beamlets that propagate behind the drive beam inside a CWA. We assume that the witness beam charge is much smaller than the drive beam. Consequently, the wakefields generated by the beamlets are smaller compared to the drive wakefields and could be neglected. Likewise, we further neglect space charge effects owing to the small charge at play. The transverse motion of the center of mass (COM) for each longitudinal $\zeta$-slice associated with each beamlet can be written as 
\begin{align}
    \frac{\partial^2{\mathbf{r}}}{\partial t^2}=\frac{e}{\gamma m_e}\mathbf{F}_\perp(\mathbf{r},\zeta),
\end{align}
where the vector $\mathbf{r}=(x,y)^{T}$ gives the transverse position of the witness-beamlet COM in the $xy$ plane.

In a wakefield accelerator, the main contribution to the wakefield comes from a steady state process when a synchronous wave traveling with the same speed as the drive bunch is generated and interacts with the witness bunch. If one neglects slippage effects, the transverse wake potential is connected to the Lorentz force $\mathbf{F}_\perp$ acting on a witness beamlet through the simple relation
\begin{align}
    \mathbf{F}_\perp=\frac{Q}{L} \mathbf{W}_\perp.
\end{align}
Here $L$ is the length of the accelerating structure and $Q$ is the total charge of the drive bunch. 
This expression is exact for the steady state wake and is still valid if we understand $F$ as an averaged Lorentz force over a known interaction length.
Assuming that no slippage occurs along the wakefield accelerator section and $z\approx c t$, we arrive at the final equation of the transverse motion in the form
\begin{align}
\label{eq:eqtm}
    \frac{\partial^2{\mathbf{r}}}{\partial z^2}=\frac{eQ}{\gamma m_e c^2 L}\mathbf{W}_\perp(\mathbf{r},\zeta).
\end{align}
\begin {figure*}
 \centering
\includegraphics[scale=0.45]{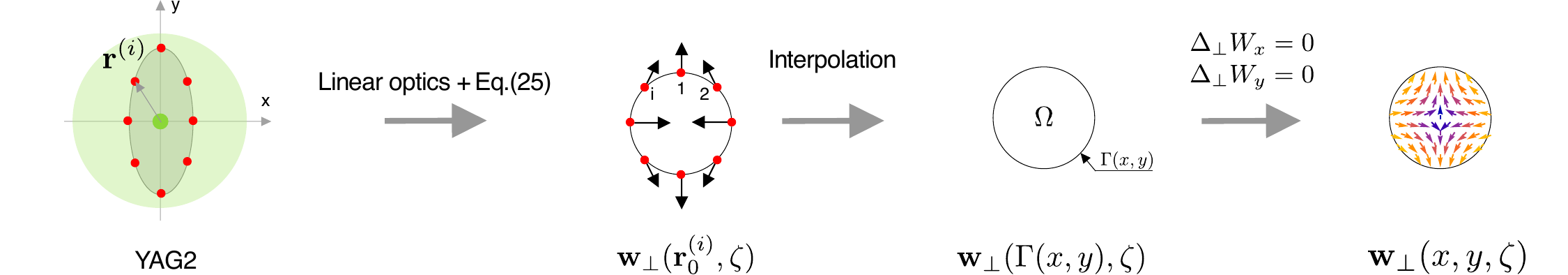}\\
\caption{Illustration of the wakepotential reconstruction algorithm: witness beamlets are displaced in the wakefield of the drive beam, defining the boundary conditions for Laplace equation on the wakepotential inside the beamlet contour.}
\label{Fig:recalg}
\end{figure*}
We introduce a parameter
\begin{align}
\label{eq:alpar}
    \alpha=\frac{eQ\max|\mathbf{W}_\perp(\mathbf{r},\zeta)|}{\gamma m_e c^2}
\end{align}
and notice that for most cases the condition $\alpha\ll1$ is fulfilled. Calculations of this parameter for two potential wakefield-acceleration experiments are presented in Appendix \ref{app:alestim}.

We then introduce the normalized transverse wake potential
\begin{align}
    \mathbf{w}_\perp(\mathbf{r},\zeta)=\frac{\mathbf{W}_\perp(\mathbf{r},\zeta)}{\max|\mathbf{W}_\perp(\mathbf{r},\zeta)|}
\end{align}
and rewrite Eq.\eqref{eq:eqtm} as
\begin{align}
\label{eq:eqtrf}
     \frac{\partial^2{\mathbf{r}}}{\partial z^2}=\alpha \frac{\mathbf{w}_\perp(\mathbf{r},\zeta)}{L}.
\end{align}
With the condition $\alpha\ll1$ Eq.\eqref{eq:eqtrf} could be solved using a perturbation series approach. We represent $\mathbf{r}$ as 
\begin{align}
\label{eq:rs}
    \mathbf{r}\approx\mathbf{r}^{(0)}+\alpha\mathbf{r}^{(1)}+\alpha^2\mathbf{r}^{(2)}+...
\end{align}
substitute it into Eq.\eqref{eq:eqtrf} and expand  $\mathbf{w}_\perp(\mathbf{r},\zeta)$ in Taylor series around  $\mathbf{r}^{(0)}$  
\begin{align}
\label{eq:eqtrpert}
     &\frac{\partial^2{\mathbf{r}^{(0)}}}{\partial z^2}+\sum\limits_{n=1}^{\infty}\alpha^n\frac{\partial^2{\mathbf{r}^{(n)}}}{\partial z^2}=\\ \nonumber&\alpha \frac{\mathbf{w}_\perp(\mathbf{r}^{(0)},\zeta)}{L}+\alpha \frac{\mathrm{J}(\mathbf{r}^{(0)},\zeta)}{L}\sum\limits_{n=1}^\infty\alpha^n\mathbf{r}^{(n)}+...
\end{align}
Here $\mathrm{J}(\mathbf{r}^{(0)},\zeta)$ is the Jacobian matrix given by
\begin{align}
  \mathrm{J}(\mathbf{r}^{(0)},\zeta)=
 \begin{pmatrix}
\partial_x w_x(\mathbf{r}^{(0)},\zeta) &&\partial_y w_x(\mathbf{r}^{(0)},\zeta)\\
\partial_x w_y(\mathbf{r}^{(0)},\zeta) &&\partial_y w_y(\mathbf{r}^{(0)},\zeta)\\
\end{pmatrix}.
\end{align}

Equating common powers of $\alpha$ on the right and left-hand sides of Eq.\eqref{eq:eqtrpert} we arrive at
\begin{align}
\label{eq:eqtrpert0}
     &\frac{\partial^2{\mathbf{r}^{(0)}}}{\partial z^2}=0, \\
\label{eq:eqtrpert1}
     &\frac{\partial^2{\mathbf{r}^{(1)}}}{\partial z^2}= \frac{\mathbf{w}_\perp(\mathbf{r}^{(0)},\zeta)}{L},\\
\label{eq:eqtrpert2}     
     &\frac{\partial^2{\mathbf{r}^{(2)}}}{\partial z^2}= \frac{\mathrm{J}(\mathbf{r}^{(0)},\zeta)\mathbf{r}^{(1)}}{L}.
\end{align}
The solution to Eq.\eqref{eq:eqtrpert0} is simply
\begin{align}
\label{eq:sl0}
   \mathbf{r}^{(0)}=\mathbf{r}_0+\boldsymbol{\beta}_0 z.
\end{align}
with $\boldsymbol{\beta}_0\equiv\mathbf{\mathbf{p}_{\perp}}_0/p_z$.
If we assume that transverse motion is non-relativistic $|\boldsymbol{\beta}_0|\ll1$ then solution to Eq.\eqref{eq:eqtrpert1} could be well approximated by
\begin{align}
\label{eq:sl1}
   \mathbf{r}^{(1)}\approx\frac{\mathbf{w}_\perp(\mathbf{r}_0,\zeta) z^2}{2L}, 
\end{align}
as well as solution to Eq.\eqref{eq:eqtrpert2}
\begin{align}
\label{eq:sl2}
   \mathbf{r}^{(2)}\approx\frac{\mathrm{J}(\mathbf{r}_0,\zeta)\mathbf{w}_\perp(\mathbf{r}_0,\zeta) z^4}{4!L^2}. 
\end{align}
Combining Eq.\eqref{eq:sl0},Eq.\eqref{eq:sl1} and Eq.\eqref{eq:sl2} with Eq.\eqref{eq:rs} we finally obtain 
\begin{align}
\label{eq:fin}
   \mathbf{r}(z,\zeta)\approx \mathbf{r}_0+\boldsymbol{\beta}_0 z+&\alpha\frac{\mathbf{w}_\perp(\mathbf{r}_0,\zeta) z^2}{2L}+\\ \nonumber &\alpha^2\frac{\mathrm{J}(\mathbf{r}_0,\zeta)\mathbf{w}_\perp(\mathbf{r}_0,\zeta) z^4}{4!L^2}+\mathcal{O}\left[\alpha^3\right]. 
\end{align}
Inversion of Eq.\eqref{eq:fin} allows one to express wake potential through the vector of initial and final positions of the beamlet as
\begin{align}
\label{eq:wkpf}
   \mathbf{w}_\perp(\mathbf{r}_0,\zeta)=&\frac{2L}{\alpha z^2}\left(\mathrm{I}-\frac{\alpha z^2 \mathrm{J}(r_0,\zeta)}{12L}+\mathcal{O}\left[\alpha^2\right] \right)\cross\nonumber \\ &\left(\mathbf{r}(z,\zeta)-\mathbf{r}_0-\boldsymbol{\beta}_0 z \right).
\end{align}
Here $\mathrm{I}$ is the identity matrix. If the
norm of the Jacobian matrix verifies $z^2||\mathrm{J}||/(12 L) \leq 1$, then Eq.\eqref{eq:wkpf} simplifies to  
\begin{align}
\label{eq:wlf0}
   \mathbf{w}_\perp(\mathbf{r}_0,\zeta)=2L\frac{\mathbf{r}(z,\zeta)-\mathbf{r}_0-\boldsymbol{\beta}_0 z}{\alpha z^2}.
\end{align}
We expand $\alpha$ according to the Eq.\eqref{eq:alpar} to finally relate the wake potential experienced by a beamlet to its initial and final positions
\begin{align}
\label{eq:wlf}
   \mathbf{W}_\perp(\mathbf{r}_0,\zeta)=\frac{2\gamma m_e c^2L}{eQ}\left[\frac{\mathbf{r}(z,\zeta)-\mathbf{r}_0-\boldsymbol{\beta}_0 z}{z^2}\right].
\end{align}
We note that under approximations above only two points are necessary to calculate the transverse wake potential at the initial position $\textbf{r}_0$ of the beamlet.   

\subsection{Reconstruction algorithm}
\label{sec:recal}

In order to reconstruct transverse wake potential over a region of the transverse plane, 
we proceed with the following sequence of measurements. 
First, both drive and witness beams' transverse distribution is recorded at YAG1 and YAG2 screens, with the CWA turned off ("passive mode"). Such a measurement establishes a reference trajectory.
For the case of dielectric slab structure, for example, this is usually accomplished by retracting the slabs far away from the beam trajectory. From this measurement, the beamlets' initial transverse positions and momenta ($\boldsymbol{\beta}_{\mathrm{Y1}}$, $\boldsymbol{r}_{\mathrm{Y1}}$) are determined at the YAG1 location. The initial conditions ($\boldsymbol{\beta}_0$, $\boldsymbol{r}_0$) are calculated based on the YAG1 values ($\boldsymbol{\beta}_{\mathrm{Y1}}$, $\boldsymbol{r}_{\mathrm{Y1}}$) by propagating the beamlets through a drift of the length equal to the distance from YAG1 to the entrance of the CWA.

Next, we switch the CWA section to the active mode, and the resulting  displacement of each individual beamlet $\boldsymbol{r}(z, \zeta)$ (at its centroid, head, or tail) is measured on YAG2. 
If YAG2 is located far from the CWA exit an additional correction factor has to be incorporated into the Eq.\eqref{eq:wlf} to account for the beam divergence due to this drift. We notice that under the assumptions of the Eq.\eqref{eq:wlf} we may drop the terms of the order $\alpha^2$ and higher in the Eq.\eqref{eq:fin}. In this case  $\boldsymbol{\beta}(z)$ inside the CWA is approximately given by
\begin{align}
    \boldsymbol{\beta}(z)\approx \boldsymbol{\beta}_0+\alpha\frac{\mathbf{w}_\perp(\mathbf{r}_0,\zeta)z}{L}+\mathcal{O}\left[\alpha^2\right].
\end{align}
Therefore, position of the beamslets on a YAG2 located at a distance $L_{\mathrm{Y2}}$ from the CWA exit could be found from
\begin{align}
\label{eq:finY}
   \mathbf{r}(L+L_{\mathrm{Y2}},\zeta)\approx \mathbf{r}_0&+\boldsymbol{\beta}_0 (L+L_{\mathrm{Y2}})\nonumber\\&+\alpha\mathbf{w}_\perp(\mathbf{r}_0,\zeta)\frac{L+2L_{\mathrm{Y2}}}{2}. 
\end{align}
Reversing Eq.\eqref{eq:finY} with respect to the wake potential we arrive at
\begin{align}
\label{eq:wlfY}
   \mathbf{W}_\perp(\mathbf{r}_0,\zeta)&=\\ \nonumber &\frac{2\gamma m_e c^2}{eQ}\left[\frac{\mathbf{r}(L+L_{\mathrm{Y2}},\zeta)-\mathbf{r}_0-\boldsymbol{\beta}_0 (L+L_{\mathrm{Y2}})}{L+2L_{\mathrm{Y2}}}\right].
\end{align}

Once Eq.\eqref{eq:wlfY} is evaluated, the wakefield is split into its two orthogonal components $W_x$ and $W_y$, and each component is interpolated on the contour $\Gamma$ using each beamlet as a mesh point on this contour. Interpolation functions are then used to solve the Laplace equations $\Delta_\perp W_{x,y} = 0$ in the region $\Omega$ enclosed by the contour $\Gamma$. 
This procedure is schematically shown in Fig.~\ref{Fig:recalg}. After initial orbit measurements have been performed, the method is able to provide integrated transverse wakefield kick measurement in a single shot.

It is worth noting that if the longitudinal profile of the beamlet is known then a change in the intensity as well as the change in the projected beam profile at the YAG2 between active and passive mode of the CWA section allows tracking not only one but several points of the beamlet that correspond to different longitudinal slices of the beamlet. This enables 3D mapping by resolving 2D maps of the transverse wakefield at several positions in $\zeta$ simultaneously. We would like to highlight that the method requires only one YAG2 image after the CWA section for the reconstruction process and thus could be considered as a single shot, beam-based method.

\subsection{Reconstruction in the ideal case}
\label{sec:simEx}
\begin{figure}[b]
    \centering
    \includegraphics[width=0.99\linewidth]{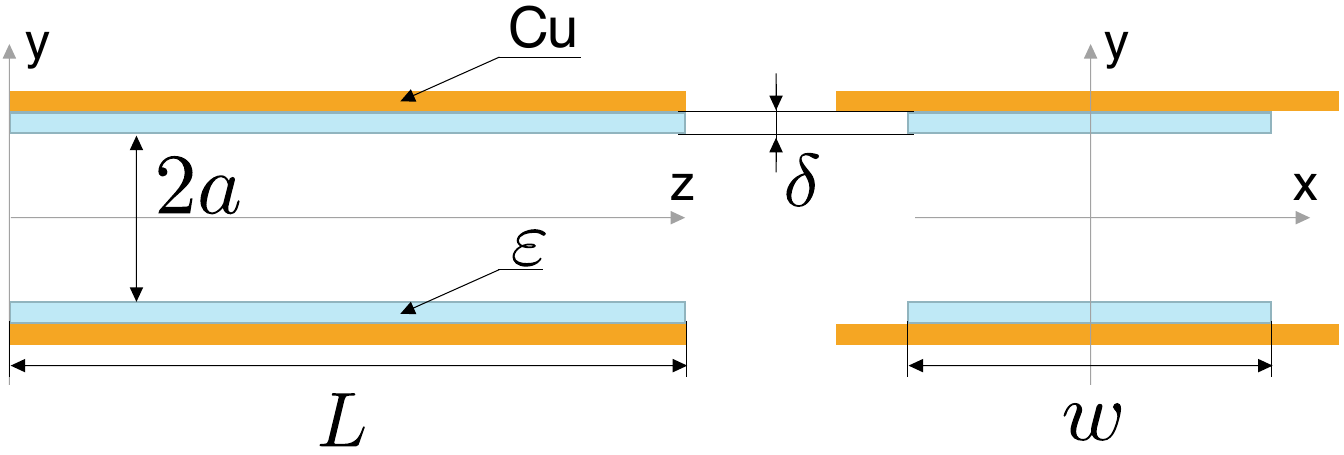}
    \caption{Schematic diagram of the slab dielectric structure.}
    \label{Fig:4}
\end{figure}
In order to validate the reconstruction algorithm, we consider point-like drive and witness bunches injected into the  dielectric slab structure (Fig.~\ref{Fig:4}), with parameters listed in Table \ref{tb:1}. This structure was recently used in the high transformer ratio experiment at the AWA facility \cite{PhysRevLett.120.114801}. A realistic start-to-end simulation is provided in Section \ref{sec:sims}.

\begin {table}[h]
\caption{Parameters of the structure and drive beam for the toy model reconstruction}
\label{tb:1}
\begin{ruledtabular}
\begin {tabular}{c c c c c c c}
$E$&$Q_{dr}$&$2a$&$\delta$&$w$&$L$&$\varepsilon$ \\
\colrule
48~MeV &2~nC&2.5~mm&150~$\mu$m&1.27~cm&15~cm&3.75 
\end {tabular}
\end{ruledtabular}
\end{table} 

\begin{figure}[hhhh!!!!!]
    \centering

    \includegraphics[width=0.99\linewidth]{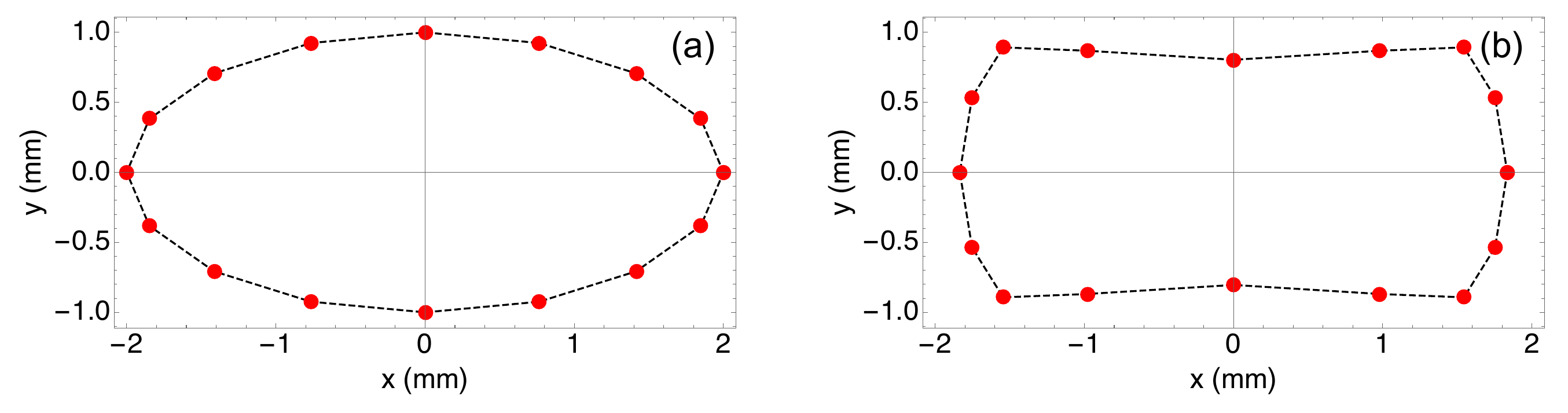}
   \includegraphics[width=0.99\linewidth]{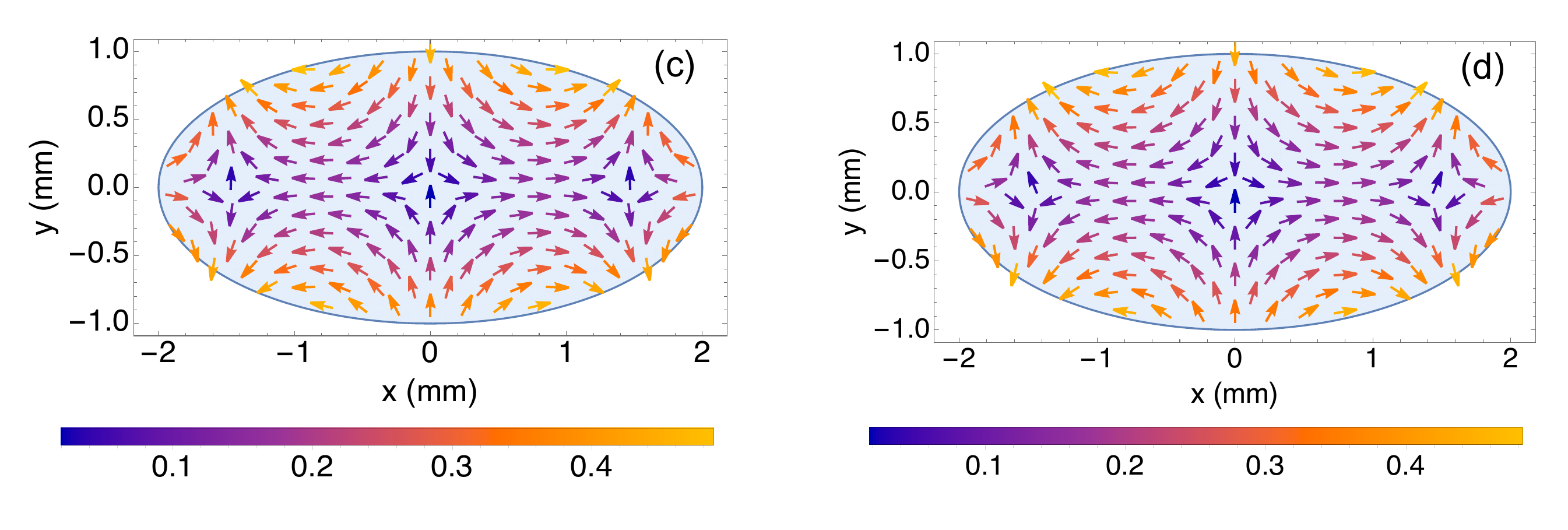}
    \caption{(a) Beamlet configuration at the structure entrance; (b) beamlet configuration at the structure exit; (c) exact transverse wakefield ($\mathbf{F}=\mathbf{W}_\perp Q/L$) inside the region surrounded by the beamlets in MV/m; (d) reconstructed transverse wakefield in MV/m.}
    \label{fig:sim}
\end{figure}

We assume that $\boldsymbol{\beta}_0=0$ for all witness beamlets, and we take the initial beam configuration at the structure entrance to be an ellipse shown in Fig.~\ref{fig:sim}(a). To calculate the wakefield generated by the drive bunch, we follow the method detailed in Ref.\cite{Baturin:slab}. Next, the evolution of the beamlets in the wakefield of the driver was tracked numerically. 

The beamlets distribution downstream of the structure is shown in Fig.~\ref{fig:sim}(b). The distance between the drive and witness pair was chosen to be $3.2$ mm. Figure~\ref{fig:sim} indicates that the beamlets distribution transforms significantly between YAG1 and YAG2 locations. The characteristic small  parameter $\alpha$ defined in Eq.\eqref{eq:alpar} is evaluated to be $\alpha\leq 5\times10^{-2}$ (see appendix \ref{app:alestim} for details). Such a small value confirms that the requirements for  the reconstruction to be valid are met (i.e. $\alpha\ll 1$).    

\begin{figure*}
    \centering
    \includegraphics[width=1\linewidth]{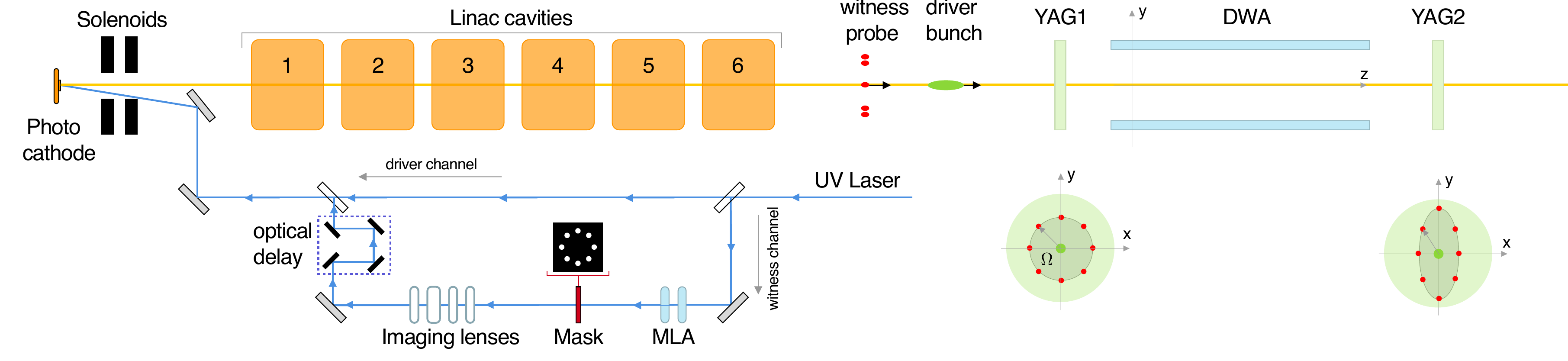}
    \caption{Schematics of the simulated AWA beam line: initial UV laser pulse is shaped transversely in the microlens array (MLA) and longitudinally in the split-and-delay optical beamline; photoemitted drive and witness beams are propagated through the dielectric wakefield accelerator (DWA). The spatial e-beam distribution is observed at three YAG viewers upstream/downstream of the DWA respectively (the schematics is not to scale). }
    \label{Fig:1}
\end{figure*}
Figure~\ref{fig:sim} compares the reconstructed and exact transverse wake potentials inside the elliptical region of interest, using the methodology from Sec.\ref{sec:recal} and Fig.~\ref{Fig:recalg}.  The comparison demonstrates a very good agreement. The reconstructed map not only captures the intricate transverse-wake-potential structure at $\zeta=3.2$ mm behind the driver but also reproduces its amplitude with a high accuracy. We conclude that the reconstruction method is consistent and produces reliable results, in the ideal case for the feasible set of experimental parameters.  
\begin{figure}[b]
    \centering
        \includegraphics[width=0.99\linewidth]{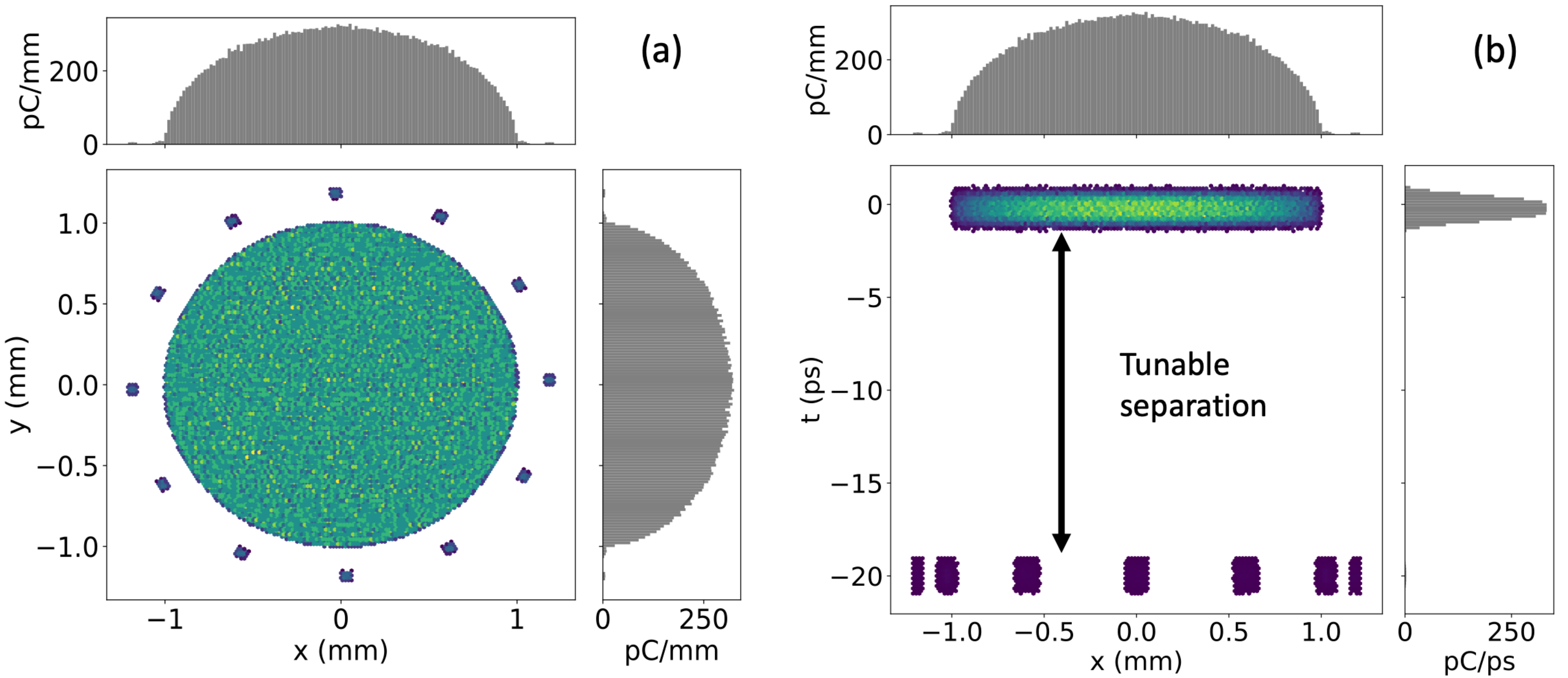}
    \caption{$xy$- (a) and $xt$- (b) projections of the final shaped laser beam phase-space at the photocathode location with a tunable separation.
    }
    \label{fig:cathode}
\end{figure}

\section{Start-to-end simulations of drive- and hollow witness bunches\label{sec:sims}}

In this section we discuss the drive-witness pair generation and present numerical simulations supporting the technique. As an example, we consider the AWA facility where proof-of-principle experiments are being considered.  For the CWA, we consider dielectric wakefield acceleration (DWA) section that previously was a subject of active study at AWA~\cite{PhysRevLett.120.114801, PhysRevAccelBeams.21.062801}. 

\subsection{Laser-based generation of hollow witness bunch}

In a photoinjector, the emitted electron-beam distribution mirrors the laser-intensity distribution and depends on the photocathode performances. Most of the cathode requires an ultraviolet (UV) laser pulse produced via a non-linear frequency up-conversion process. There has been significant research effort towards arbitrary transverse laser shaping~\cite{PhysRevAccelBeams.20.103404, Halavanau:2018ajm, Halavanau:2019kwc} in support of wakefield experiments~\cite{PhysRevLett.120.114801, PhysRevLett.124.044802, ROUSSEL2018130}. 
Generation of witness and drive bunches for the suggested scheme is relatively straightforward. An example of implementation is diagrammed in Fig.~\ref{Fig:1}: a UV pulse produced by the laser system is split along two optical lines and then recombined. One of the optical lines includes a variable-delay stage to control the temporal separation between the drive pulse and witness pulse. The delay line incorporates the transverse-shaping optical elements that generate a necklace-like hollow beam. We consider using a microlens array to produce a homogenized distribution followed by a mask and necessary optics to image the hollow pattern on the cathode surface.
A hollow transverse UV laser profile can also be obtained with digital micro-mirror devices \cite{PhysRevAccelBeams.20.080704,PhysRevSTAB.18.023401}, axicon lenses or by employing Laguerre-Gaussian laser modes~\cite{Maurer_2007}.

The first mention of hollow electron beams was associated with electron ring accelerator (ERA) project, e.g. see Refs \cite{KEEFE1971541,Keefe1976,LAWSON1968,Pellegrini1969}. Hollow beams were also considered as wakefield \textit{drive} beams in the Resonance Wakefield Transformer (RWT) collider proposal \cite{Weiland:1985vg, Holtkamp:1990bw, Bialowons:1986gus, Decker:1988qb, Bialowons:1987sc, PhysRevSTAB.12.051301}.
That work pointed out challenges in creating and propagating a stable hollow electron beam, specifically due to negative mass instability and resistive wall wakefields. However, in the case of a necklace-like hollow beam consisting of small round beamlets, these instabilities are suppressed. The necklace-like pattern also allows us to "tag" individual beamlets (and its evolution with and without being affected by the transverse wakefield).

The macroparticle distributions for both drive and witness beams were generated using {\sc distgen} package \cite{distgen}. Figure~\ref{fig:cathode} illustrates the spatial and temporal structure of the drive-witness pair.
In the simulations, the drive bunch has a bunch charge of 0.5 nC, while the necklace witness bunch consists of twelve 300-fC beamlets yielding a total charge of 3.6 pC. Both bunches were generated using a laser pulse with a 2.5-ps (FWHM) flat-top temporal distribution. The delay between the drive and witness  bunches is variable with the smallest attainable delay ultimately limited by the Coulomb field associated with the drive bunch which could significantly affect the dynamics of the witness bunch.

\subsection{Beam dynamics in the AWA photoinjector}

AWA photoinjector incorporates a normal-conducting L-band electron gun with Cs$_2$:Te photocathode and six 1.3-GHz accelerating cavities boosting the beam energy up to 72 MeV~\cite{Power:2010zza}; see Fig.~\ref{Fig:1}. The RF gun is nested in three solenoidal lenses to control the emittance-compensation process. Additionally, three solenoidal lenses are located downstream of cavities 1, 3, and 5.

\begin{figure}[b]
    \centering
    \includegraphics[width=0.99\linewidth]{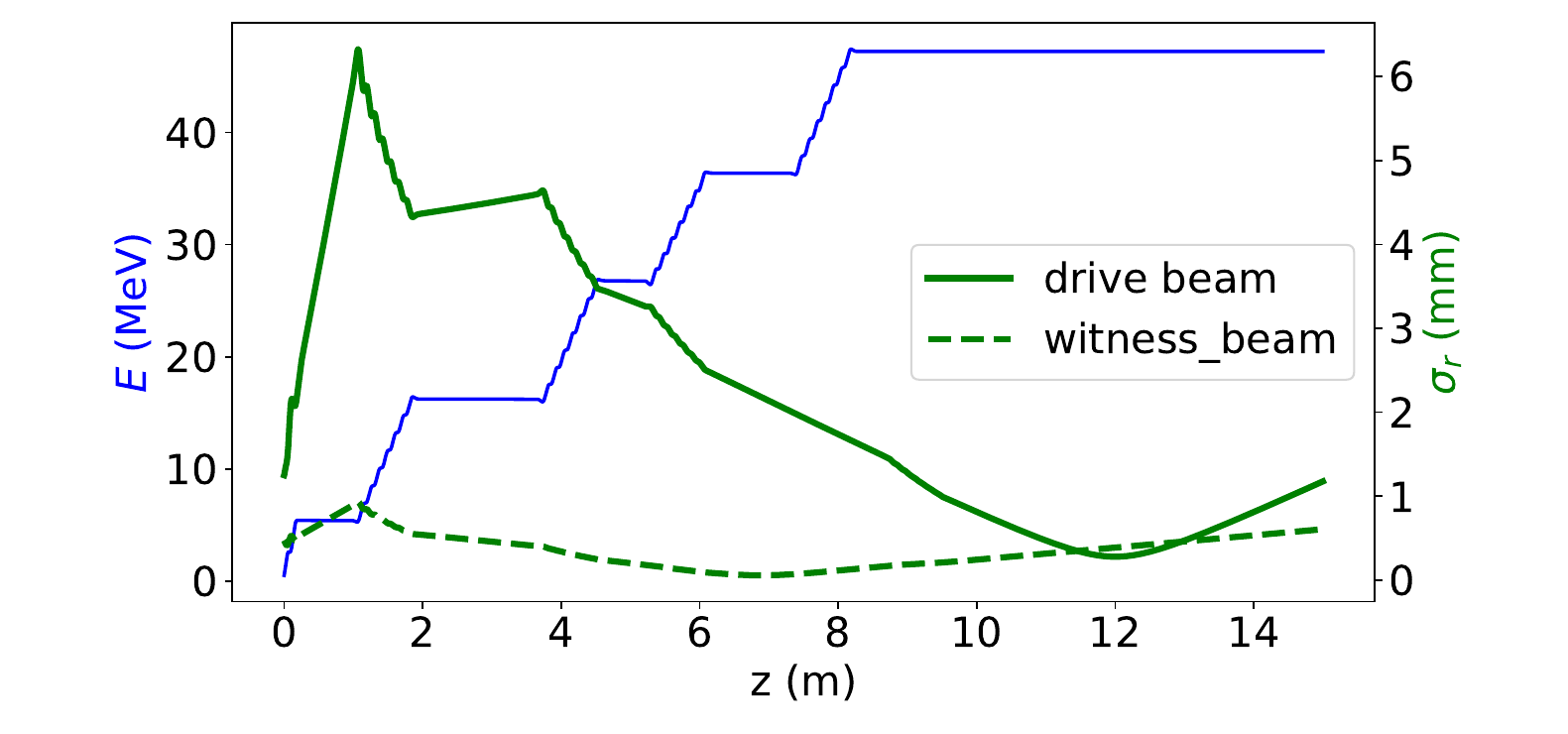}
    \caption{Electron beam energy and RMS spot sizes associated with the drive (solid trace) and witness (dashed trace) beams as a function of distance $z$ in AWA photoinjector. The DWA structure is located at $z=12$ m.}
    \label{awa:envelope}
\end{figure}

\begin{figure}[t]
    \centering
    \includegraphics[width=0.99\linewidth]{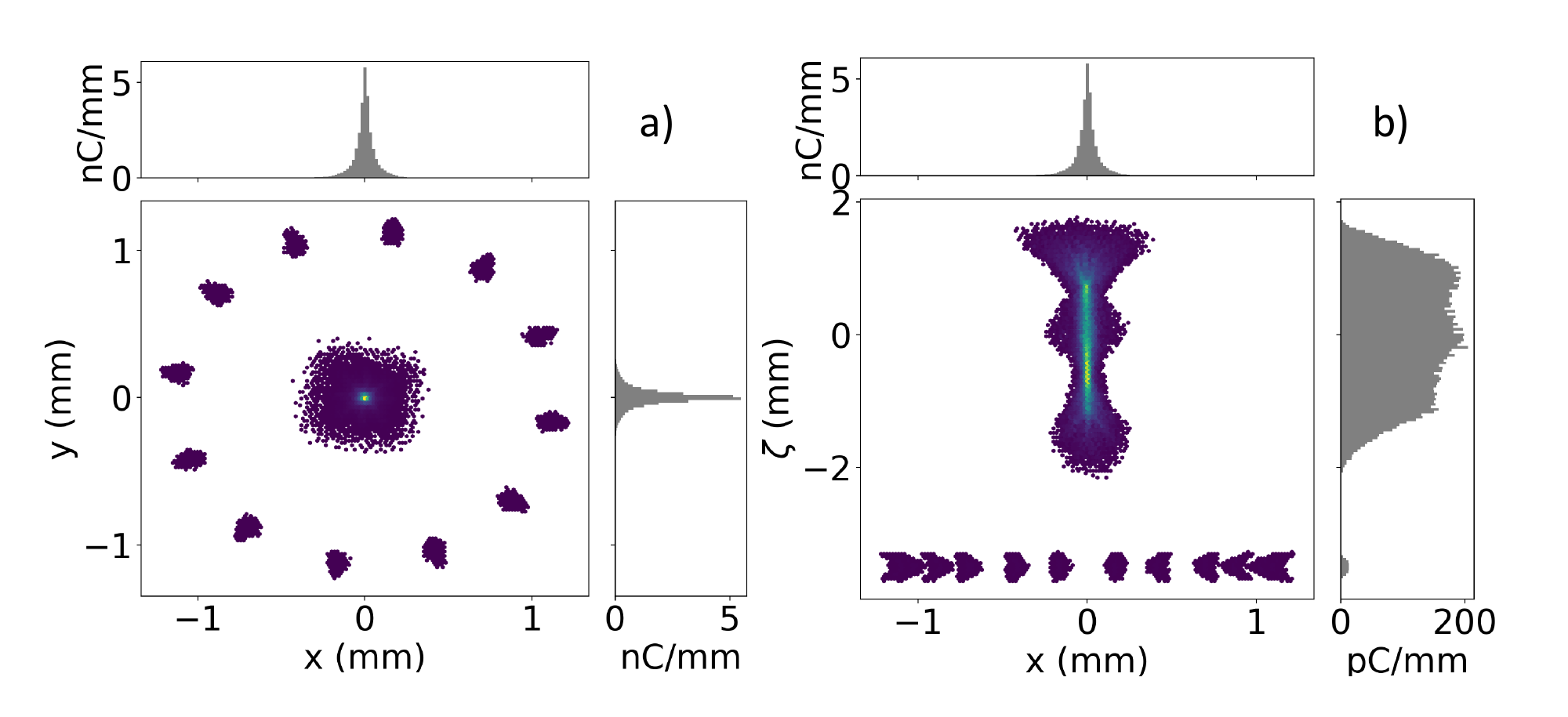}
    \caption{$xy$- (a) and $x\zeta$- (b) projections of electron beam phase-space at the entrance of the DWA.
    }
    \label{fig:dwa}
\end{figure}

The numerical model of AWA beamline was implemented in the {\sc impact-t} beam-physics program~\cite{Qiang:2006wq,PhysRevSTAB.9.044204} which has been extensively benchmarked against other programs; see Refs.~\cite{Neveu:2017uzs,Mayes:2018pph}. The {\sc impact-T} program includes a three-dimensional quasi-static space-charge algorithm where the Poisson equation is solved in the bunch's average rest frame. An example of optimized settings showcasing the evolution of the drive and witness bunches envelopes appears in Fig.~\ref{awa:envelope}. For these simulations, the drive and witness bunch are tracked individually, in order to avoid using a large number of longitudinal bins in the space charge algorithm. Both drive and witness bunch lie within the same 1.3-GHz RF-bucket with variable separation of up to 8~mm in the simulations. Transverse beam optics is globally optimized to allow for a beam waist at the entrance of the DWA section, spatially resolving the drive and witness beams. 
We consider the wakefield interaction to occur at a distance $z=12$~m from the photocathode. The corresponding drive and witness bunch distributions at this location are summarized in Fig.~\ref{fig:dwa} and confirm that the necklace-like witness bunch structure is preserved albeit for some radial smearing. Reducing the bunch charge does mitigate this effect but a very low charge makes it difficult to register the beamlet pattern on the YAG screen. Increasing the witness-beam charge, on the contrary, may lead to multiple transverse instabilities and breaks the axial symmetry of the beamlet arrangement~\cite{Halavanau:IPAC20}. It should be noted that the reconstruction algorithm is insensitive to such type of beamlet distortion.  

While traversing the dielectric structure, the beamlets' size remains relatively constant, and therefore we apply our technique to their center of mass. Comparing Fig.~\ref{fig:cathode} with Fig.~\ref{fig:dwa} one can note the beamlet distribution on the cathode is translated to DWA entrance almost identically. Since both drive and witness beams share the same beam optics and have the same beam energy, the only parameter to effectively control the witness beam's phase advance is its size on the cathode. We found, via numerical simulations, that it is possible to image the witness beam cathode image at a given location $z$, while focus the drive beam at the same location.

\section{Simulations of the wakefield-mapping technique}

\begin {table}[b]
\caption{Parameters of the structure and drive beam for the start to end simulation}
\label{tb:1a}
\begin{ruledtabular}
\begin {tabular}{c c c c c c c}
$E$&$Q_{dr}$&$2a$&$\delta$&$w$&$L$&$\varepsilon$ \\
\colrule
48~MeV &0.5~nC&2.5~mm&150~$\mu$m&1.27~cm&15~cm&3.75 
\end {tabular}
\end{ruledtabular}
\end{table} 

\begin{figure}[b]
    \centering
    \includegraphics[width=1\linewidth]{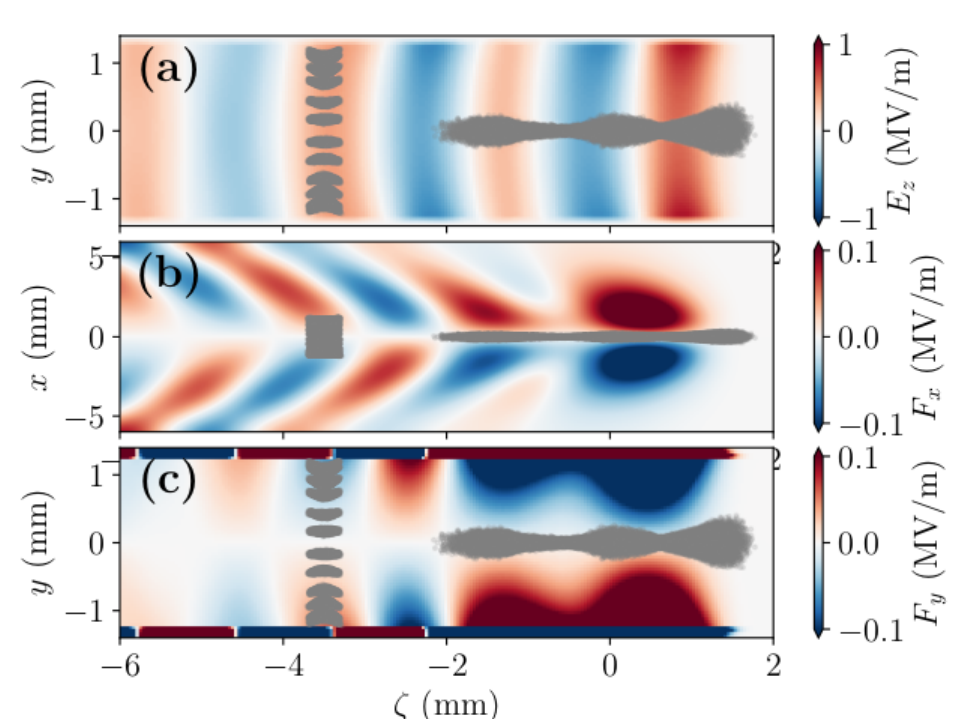}
    \caption{Snapshot of the beam distribution (grey dots) with superimposed wakefields (colormap) simulated with {\sc warp} at the end of the DWA structure. For this example the witness bunch is located at $\zeta=3.5$~mm behind the drive bunch. The longitudinal (a), horizontal (b) and vertical (c) wakefields are shown in respectively $(\zeta,y$), $(\zeta,x)$, and $(\zeta,y$). For plot (b) and (c) the scale was forced to the interval $[-0.1, 0.1]$~MV/m so that higher-field values (i.e. in the vicinity of the drive bunch and in the dielectric material) appear as saturated.  
    \label{Fig:w1}}
\end{figure}
The realistic beam distributions simulated in the previous section were used to demonstrate the application of the proposed technique in the close to actual experimental environment. The beam generated as described in the previous section (Fig.~\ref{fig:dwa}) was injected in the DWA with parameters listed in Tab.~\ref{tb:1a} and the beam-structure interaction was modeled with the particle-in-cell (PIC) finite-different time-domain (FDTD) program {\sc warp}~\cite{Vay_2012}. In {\sc warp}, we performed three-dimensional simulations of the interaction. We reduce the size of the computational domain by implementing a moving-window approach where the fields are solved within a window co-moving with the beam. The moving-window length was set to 40-mm along the direction of propagation while its transverse dimension were set to accommodate the physical transverse boundaries of the DWA (i.e. $w=12.7$~mm and $2(a+\delta)=2.8$~mm in the horizontal and vertical dimensions respectively). Perfect electrical conductor (PEC) boundaries were applied in the $x=\pm w/2$ and $y=\pm (a+\delta)$ planes while the boundaries in the plane orthogonal to $\hat{z}$ were set to a perfectly matched layer (PML) so to avoid reflection of the fields. The computational mesh was selected to approximately realize a cell size on the order of $(40$~\textmu{m}$)^3$. The drive and witness bunches simulated with {\sc impact-t} are directly imported in the {\sc warp} simulation and the associated electromagnetic fields initialized at the first iteration. To avoid effects associated with the edge transition, the bunches are started within an arbitrarily-long DWA, and the simulation is performed until the witness has interacted with the drive bunch over a length of 15~cm.

\begin{figure}[t]
    \centering
    \includegraphics[width=0.99\linewidth]{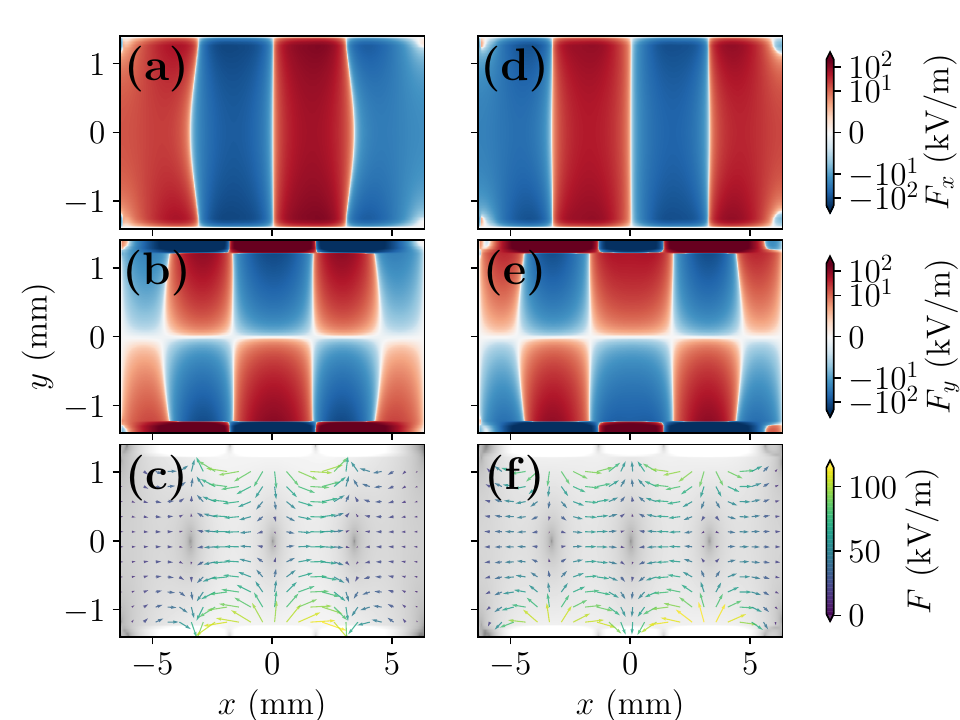}
    \caption{Horizontal (a,d), vertical (b,e), and total transverse (c,f) transverse wakefield computed in the transverse $(x,y)$ plane for a witness-bunch delay of 2.5 (a-c) and 3.5~mm (d-f). In plots (c) and (f), the superimposed gray color map represents the modulus of the wakefield (with darker shade corresponding to lower values).}
    \label{Fig:w2}
\end{figure}

Figure~\ref{Fig:w1} displays a snapshot of the beam distribution and associated longitudinal field and transverse forces at the end of the DWA. Within the drive-bunch region ($\zeta \in [-2,2]$~mm), the transverse fields [which account for both the radiative (wakefield) and velocity (space-charge) fields] are dominated by space charge. The coordinates of the macroparticle ensemble representing the witness bunch were  propagated through a drift of $L_{Y2}=0.865$~m length downstream of the DWA (using a simple linear transformation) and finally plugged into the transverse-wakefield reconstruction algorithm. We used {\sc warp} to compute the wakefield behind the drive bunch without the presence of the witness bunch. The corresponding transverse wakefields simulated at $\zeta=2.5$ and $3.5$~mm appear in Fig.~\ref{Fig:w2}. These wakefields are computed as $\pmb F(x,y)=(E_x-cB_y, E_y+cB_x)^T$ where $\{\pmb E, \pmb B\}(x,y)$ are the electromagnetic fields simulated with {\sc warp} and interpolated on a plane $(x,y)$ for the two cases of delays. Despite the relatively-simple geometry of the DWA, the transverse wakefield has a rich structure [see Figs.~\ref{Fig:w2}(c,f) and Figs.~\ref{Fig:4}(c,d)] which remains to be  investigated experimentally.

\begin{figure}[thb]
    \centering
    \includegraphics[width=0.99\linewidth]{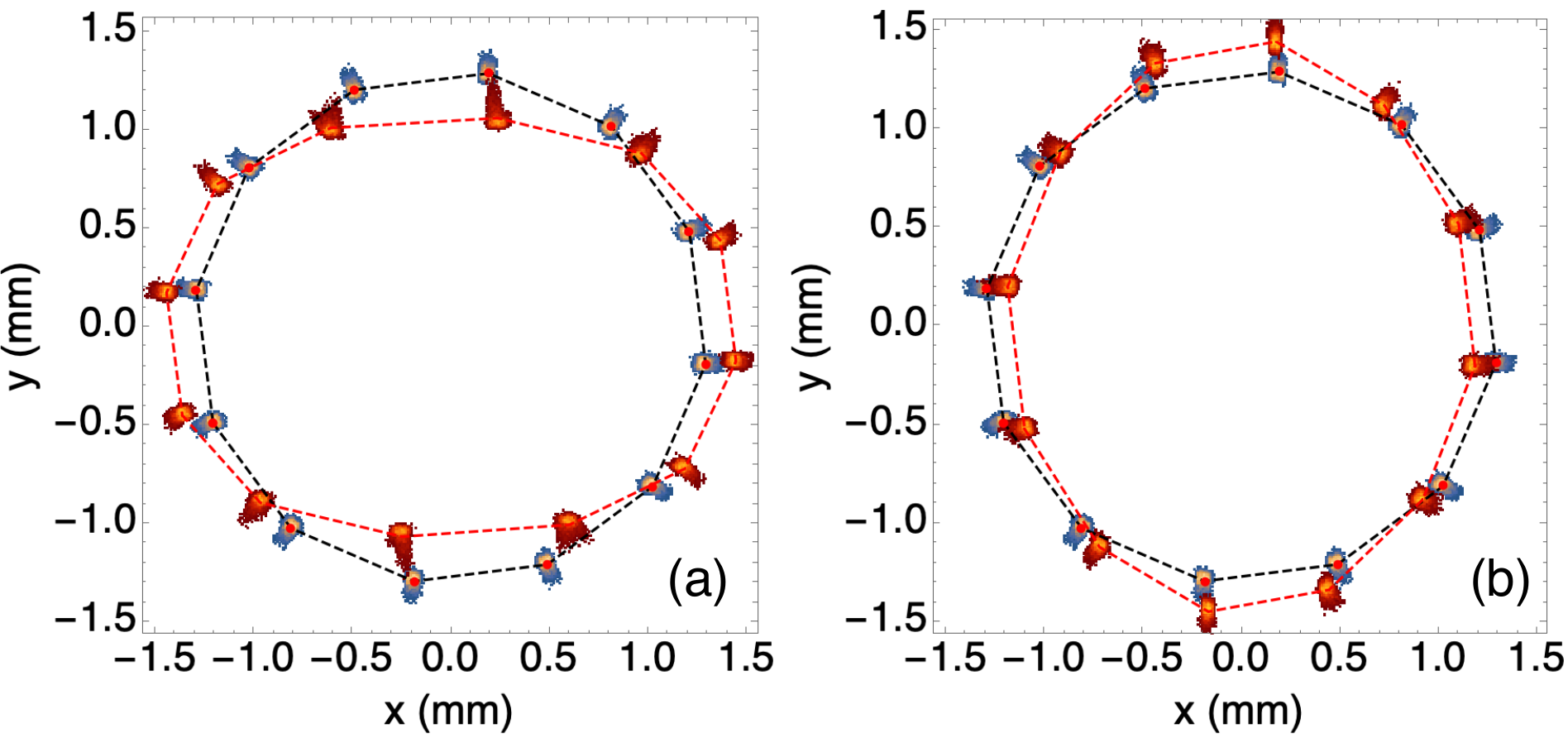}
    \caption{$xy$- projections of beamlets phase-space 0.865 m downstream of the dielectric insert exit (YAG2 location). Blue histograms represent beamlets without DWA section and red histograms represent results with the DWA. Panel a) correspond to a 2.5=mm separation between the longitudinal driver centroid and the longitudinal beamlet ring centroid, panel b) corresponds to a 3.5-mm separation between the longitudinal driver centroid and the longitudinal beamlet ring centroid. Dashed lines represent contour shapes in both cases (blue without the DWA and red with the DWA).  
    }
    \label{fig:9}
\end{figure}

\begin{figure}[t]
    \centering
    \includegraphics[width=0.99\linewidth]{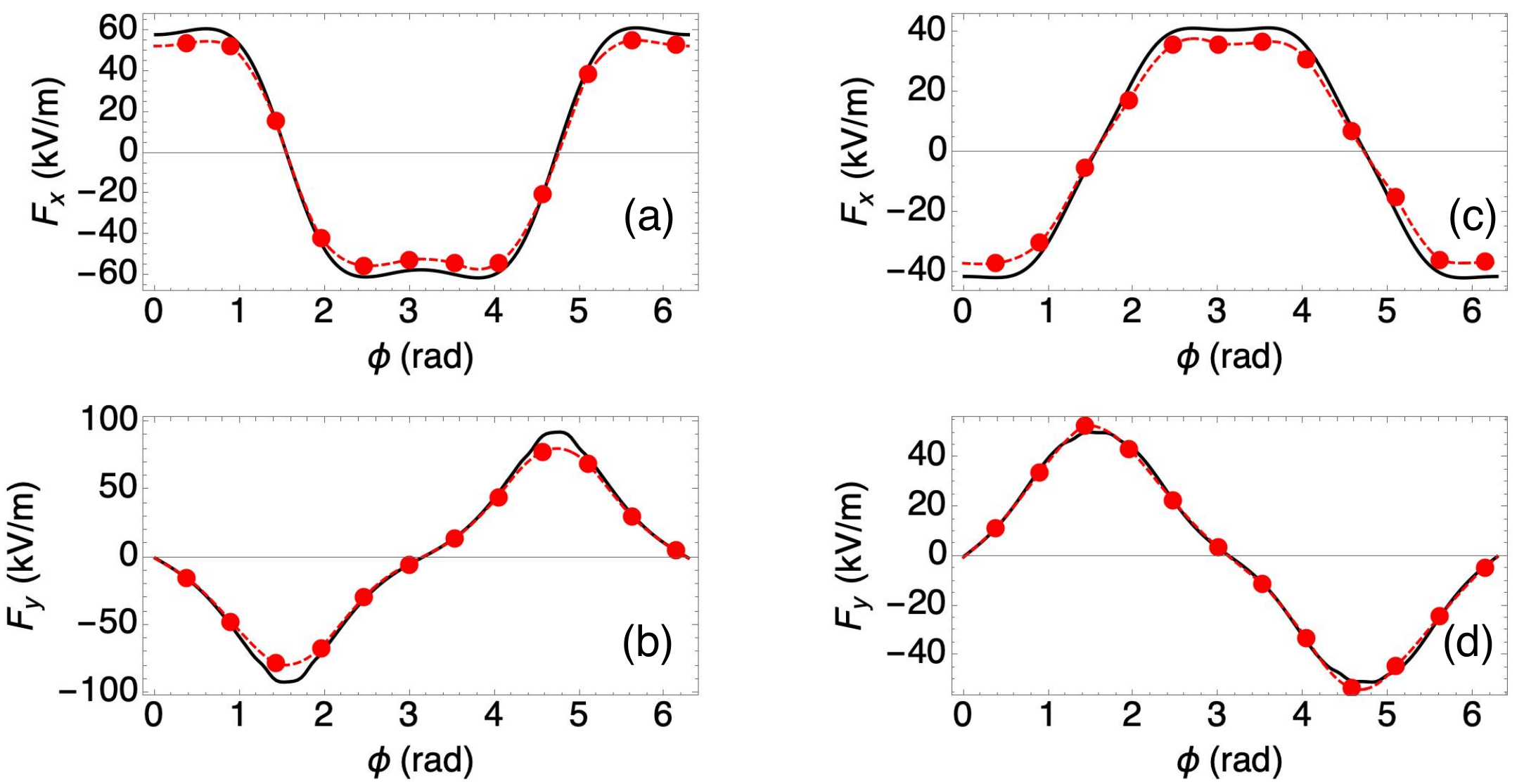}
    \caption{Horizontal and vertical components of the transverse wakefield on a circle of a radius 1.13~mm (average radius of the beamlet ring without DWA)  as a function of the polar angle $\phi$. The black line is for the exact wakefield extracted from the {\sc warp} simulation without the presence of the witness bunch; red dots result of the reconstruction algorithm applied to the {\sc warp} output; red dashed line - fourth-order spline interpolation of the data presented by points. Panel a) and panel b) correspond to the 2.5-mm separation between the longitudinal driver centroid and the longitudinal beamlet ring centroid; panel c) and panel d) correspond to the 3.5 mm separation between the longitudinal driver centroid and the longitudinal beamlet ring centroid }
    \label{fig:10}
\end{figure}

Results of the beamlet deformations for the two considered separations (of 2.5~mm and 3.5~mm) between drive and witness centroids appear in Fig.~\ref{fig:9}. To proceed further, we first identify the projected center of mass (COM) for each beamlet. Next, assuming that the projected COM corresponds to the longitudinal COM, we apply Eq.\eqref{eq:wlfY} and reconstruct transverse wakefield $\mathbf{F}=\mathbf{W}_\perp Q/L$ on the initial beamlet contour of 1.13~mm radius at the DWA entrance. In Fig.~\ref{fig:10} interpolation results along with the reconstructed values are compared with the wakefield extracted from the {\sc warp} simulation as described above (the simulations were done without the presence of the witness bunch and electromagnetic fields were calculated at $\zeta=2.5$ and $3.5$ mm delay from the driver beam longitudinal COM). 

Figure~\ref{fig:10} indicates that reconstruction provides quite accurate results and reproduces the amplitude of the wakefield with a reasonable tolerance. With a modest number of beamlets, the shapes of the curves (values of the wakefield on a circular contour) presented in Fig.~\ref{fig:10} are closely captured.

\begin{figure}[t]
    \centering
    \includegraphics[width=0.99\linewidth]{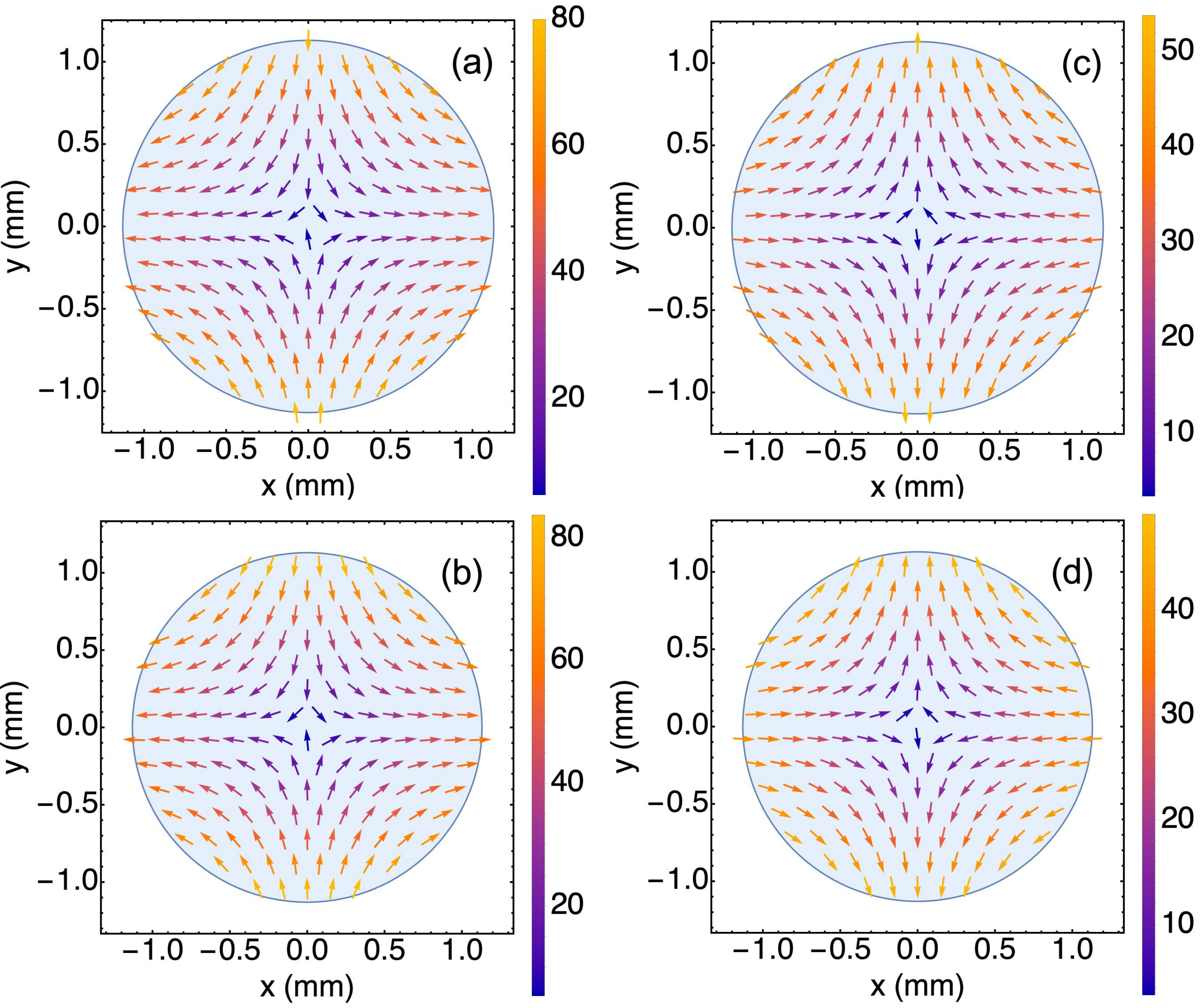}
    \caption{Reconstructed - panel a) and extracted from a separate {\sc warp} simulation - panel b) transverse wakefield ($\mathbf{F}=\mathbf{W}_\perp Q/L$) in kV/m for the 2.5 mm separation between the longitudinal driver centroid and longitudinal beamlet ring centroid. Reconstructed - panel c) and extracted from a separate {\sc warp} simulation panel d) transverse wakefield in kV/m ($\mathbf{F}=\mathbf{W}_\perp Q/L$) for the 3.5 mm separation between the longitudinal driver centroid and the longitudinal beamlet ring centroid.  
    }
    \label{fig:11}
\end{figure}

We finalize the comparison in Fig.~\ref{fig:11}. We present reconstructed and exact field maps of the transverse wakefield extracted from a separate {\sc warp} simulation.
We observe that in the two considered cases, the method captures the structure of the transverse wake potential with high accuracy. However, there is a slight disagreement in the amplitude that could be attributed to the effects of the beamlets' longitudinal deformation and inaccuracy of the centroid tracking. A possible method to improve the accuracy would be to introduce a $r$-$t$ correlation within the beamlets to encode the positions of the longitudinal slices on the $xy$ projection. Consequent tracking of the individual slices may enhance the resolution as well as enable full 3D mapping.

We point out a limitation to our technique's accuracy associated with the finite length of the beamlet. Ultimately the smaller is the ratio of the beamlet length to the transverse wakefield longitudinal variation, the higher is the resolution. This, as well as individual longitudinal slice tracking of the beamlet, is a subject of future studies. Another factor that may limit the mapping's accuracy is a weak wakefield force that will result in indistinguishable or very small displacements of the beamlet on the downstream YAG. While it is important for the diagnostics of regular accelerator components, it is not the case for the structure-based and plasma-based wakefield accelerators where the longitudinal and the transverse wakefields are known to be very large \cite{BBU2,BBU3,Brendan:2020}. 

\section{Conclusions}

We have proposed and demonstrated, via numerical simulations, a single-shot transverse-wakefield measurement technique.  In the presented study, we considered a case of a DWA using realistic AWA beam parameters. The method relies on a transversely-shaped, necklace-like witness bunch that samples integrated wakefield kick over a closed contour. In simulations, we verified that the required witness bunch could indeed be formed and transported through the dielectric slab at AWA facility, without significant challenges. The reconstructed wake potentials are in excellent agreement with {\sc warp} simulations.

The method does not require additional beamline diagnostics besides standard transverse beam-density monitors (e.g., scintillating screens). 
It should be noted that the presented technique is general, as the condition described by Eq.~\eqref{eq:wkpf}  is valid for an arbitrary structure with a net neutral channel and slow-varying fields within the beamlets composing the witness probe. Therefore we expect the technique to find its application in characterizing the transverse wakefields of various structures, including dielectric, corrugated and tapered waveguides, and hollow plasma channels~\cite{Gessner1, Gessner2}). With some additional modifications, it could also be adapted to the case of plasma-wakefield accelerators operating in the blowout regime. 
 
 Finally, the numerical complexity of the wakefield reconstruction algorithm could be alleviated with machine-learning tools. Such an approach could significantly reduce the time needed to map the wakefields and ultimately provides an online diagnostics for three-dimensional wakefield measurements.

\begin{acknowledgments}
A.H. is grateful to G. Stupakov, T. Raubenheimer, C. Mayes, and J. Rosenzweig for many insightful discussions; P.P. would like to thank D. Grote, R. Jambunathan, R. Lehe, and J.-L. Vay for their help with {\sc warp}. A.H. was supported by the U.S. Department of Energy (DOE) Contract No. DE-AC02-76SF00515 with SLAC, P.P. by the U.S. DOE awards No. DE-SC0018656 to Northern Illinois University and contract No. DE-AC02-06CH11357 with Argonne National Laboratory. S.S.B. would like to acknowledge the Foundation for the Advancement of Theoretical Physics and Mathematics "BASIS'' $\#$22-1-2-47-17 and ITMO Fellowship and Professorship program.
\end{acknowledgments}

\appendix
\section{Numerical estimations of the \texorpdfstring{$\alpha$}{alpha}  parameter \label{app:alestim}}

To estimate the small parameter introduced in Eq.\eqref{eq:alpar} we consider two experimental setups one that is based on AWA capabilities and a second one that is based on FACET-II capabilities and experimental DWA program at FACET-II. 

We note that the longitudinally extended bunch has lower coupling to the structure modes than the point-particle bunch with the same charge. This allows one to estimate $\max|\mathbf{W}_\perp(\mathbf{r},\zeta)|$  as
\begin{align}
\label{eq:ubA}
    \max|\mathbf{W}_\perp(\mathbf{r},\zeta)|\leq\max|\mathbf{G}_\perp(\mathbf{r},\zeta)|,
\end{align}
where $G$ is the Green's function for the structure. 
First, we consider AWA case and structure with the parameters that was used in Ref.\cite{PhysRevLett.120.114801}. For the convince we list the parameters of the structure again in Table \ref{tb:a1}.

\begin{figure}[h!]
    \centering
    \includegraphics[width=1.\linewidth]{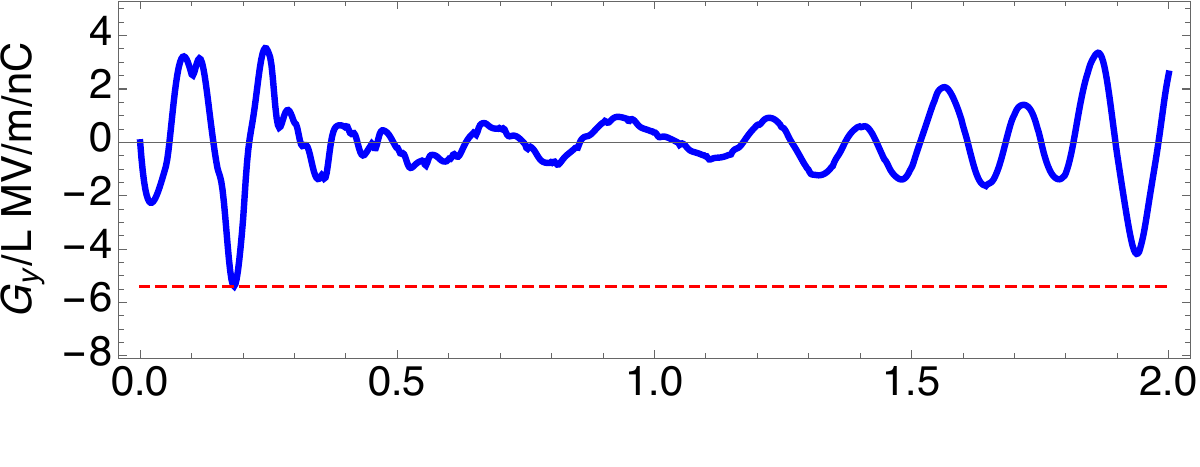}
    \includegraphics[width=1.\linewidth]{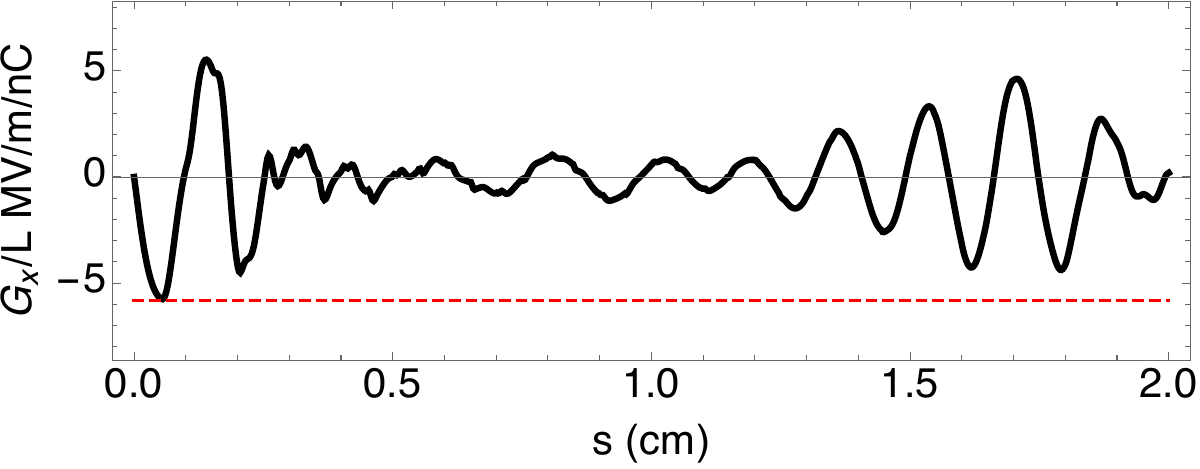}
    \caption{AWA case. Transverse Green's functions per unit length $G_x/L$ and $G_y/L$ for the driver position $x_0=0$ and $y_0=625$ $\mu$m and witness position at $x=1$ mm and $y=1$ mm. Red dashed lines indicate  $\max|G_{x,y}|$.}
    \label{fig:AWAwk}
\end{figure}

\begin{table}[h]
\caption{Parameters of the structure and driver beam for the AWA case \cite{PhysRevLett.120.114801}.}
\label{tb:a1}
\begin{ruledtabular}
\begin {tabular}{c c c c c c c}
$E$&$Q_{dr}$&$2a$&$\delta$&$w$&$L$&$\varepsilon$ \\
\colrule
48~MeV &2~nC&2.5~mm&150~$\mu$m&1.27~cm&15~cm&3.75 
\end {tabular}
\end{ruledtabular}
\end{table}
Next we consider the worst-case scenario when the driver beam is displaced from the structure center towards the dielectric and is located at $y_0=a/2$. We assume the beamlet position to be at $x=0.8a$ and $y=0.8$ where the modulus of the transverse wake potential is maximum. 
Transverse components of the wake potential per unit length $G_x/L$ and $G_y/L$ for the parameters listed in Table \ref{tb:a1} and transverse positions of the driver and witness listed above are shown in Fig.~\ref{fig:AWAwk}. 

As it could be see from Fig.~\ref{fig:AWAwk} maximum values of $x$ and $y$ components are $\max|G_x/L|\approx 5.8$~MV/m/nC and $\max|G_y/L|\approx$ 5.4 MV/m/nC. With this maximum amplitude could be estimated as $\max|\mathbf{G}_\perp|\approx 8$~MV/m/nC.
This with Eq.\eqref{eq:alpar}, Eq.\eqref{eq:ubA} and parameters from the Table \ref{tb:a1} gives
\begin{align}
    \alpha_{\mathrm{AWA}}\leq 0.05.
\end{align}

As a second example, we consider a structure that is a potential candidate for the DWA experiment at FACET-II. Parameters of the structure a listed in Table \ref{tb:a2}. As a reference parameters of a recent DWA slab experiment at FACET \cite{Brendan:2020} are extrapolated to the half-meter structure.
\begin{table}[h!]
\caption{Parameters of the structure and driver beam for the potential FACET-II case.}
\label{tb:a2}
\begin{ruledtabular}
\begin {tabular}{c c c c c c c}
$E$&$Q_{dr}$&$2a$&$\delta$&$w$&$L$&$\varepsilon$ \\
\colrule
20~GeV &3~nC&480~$\mu$m&210~$\mu$m&2.47~cm&500~cm&3.75 
\end {tabular}
\end{ruledtabular}
\end{table}

\begin{figure}[h!]
    \centering
    \includegraphics[width=1.\linewidth]{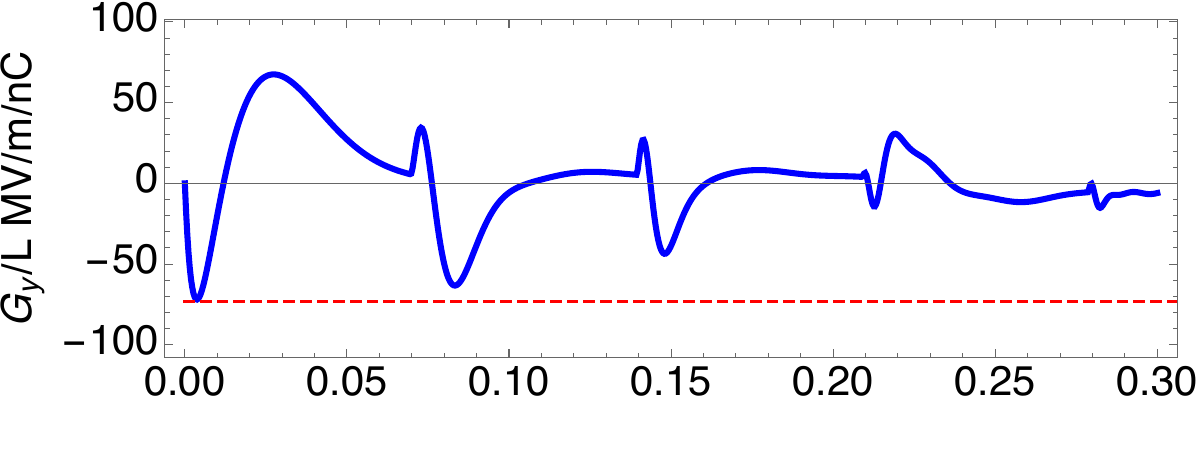}
    \includegraphics[width=1.\linewidth]{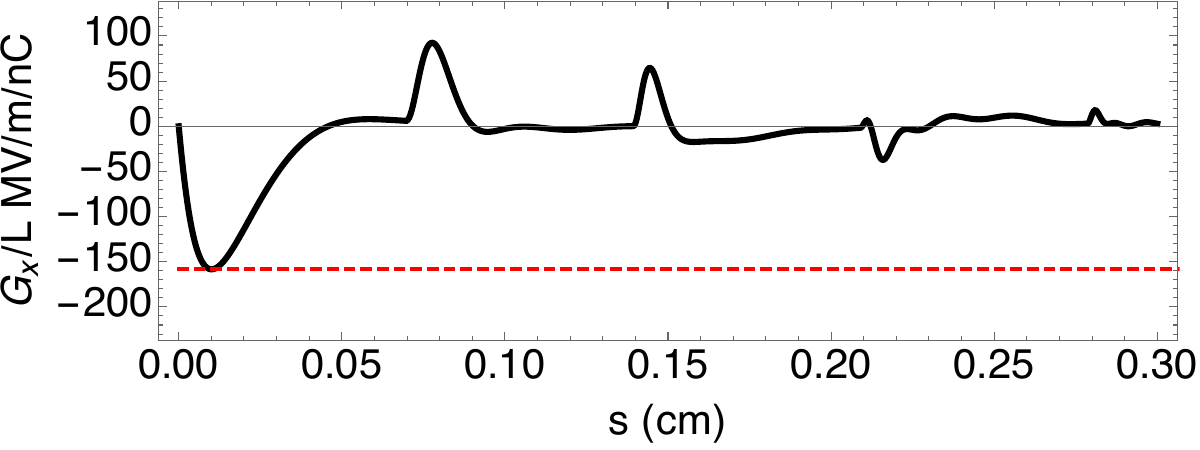}
    \caption{FACET-II case. Transverse Green's functions per unit length $G_x/L$ and $G_y/L$ for the driver position $x_0=0$ and $y_0=120$ $\mu$m and witness position at $x=192$ $\mu$m and $y=192$ $\mu$m. Red dashed lines indicate  $\max|G_{x,y}|$.}
    \label{fig:FACETwk}
\end{figure}

We again consider the worst-case scenario when the driver beam is displaced from the structure center towards the dielectric and is located at $y_0=a/2$. We assume beamlet position to be at $x=0.8a$ and $y=0.8$ where the modulus of the transverse wake potential is maximal. 
Transverse components of the wake potential per unit length $G_x/L$ and $G_y/L$ for the parameters listed in Table \ref{tb:a2} and transverse positions of the drive and witness beams listed above are shown in Fig.~\ref{fig:FACETwk}. 

As it could be see from Fig.~\ref{fig:FACETwk} maximum values of $x$ and $y$ components are $\max|G_x/L|\approx 73$~MV/m/nC and $\max|G_y/L|\approx 158$~MV/m/nC. With this maximum amplitude could be estimated as $\max|\mathbf{G}_\perp|\approx 174$~MV/m/nC.
This with Eq.\eqref{eq:alpar}, Eq.\eqref{eq:ubA} and parameters from the Table \ref{tb:a1} gives
\begin{align}
    \alpha_{\mathrm{FACET}}\leq 0.13.
\end{align}

\bibliographystyle{unsrt}
\bibliography{references}

\end{document}